**Title:** Confined Magnetization at the Sublattice-Matched Ruthenium Oxide Heterointerface

Yiyan Fan,† Qinghua Zhang,† Ting Lin, He Bai, Chuanrui Huo, Qiao Jin, Tielong Deng, Songhee Choi, Shengru Chen, Haitao Hong, Ting Cui, Qianying Wang, Dongke Rong, Chen Liu, Chen Ge, Tao Zhu, Lin Gu, Kuijuan Jin*, Jun Chen*, and Er-Jia Guo*

Y. Y. Fan, C. R. Huo, and Prof. J. Chen
Beijing Advanced Innovation Center for Materials Genome Engineering, Department of Physical Chemistry, University of Science and Technology Beijing, Beijing, 100083, China
E-mail: junchen@ustb.edu.cn

Q. H. Zhang, T. Lin, Q. Jin, S. Choi, S. R. Chen, H.T. Hong, T. Cui, Q. Y. Wang, D.K. Rong, Prof. C. Ge, Prof. T. Zhu, Prof. K.J. Jin and Prof. E-J Guo
Beijing National Laboratory for Condensed Matter Physics and Institute of Physics, Chinese Academy of Sciences, Beijing 100190, China
E-mail: kjjin@iphy.ac.cn and ejguo@iphy.ac.cn

H. Bai and Prof. T. Zhu
Spallation Neutron Source Science Center, Dongguan 523803, China

H. Bai
Institute of High Energy Physics, Chinese Academy of Sciences, Beijing 100049, China

S. R. Chen, H.T. Hong, T. Cui, Q. Y. Wang, D.K. Rong, Prof. K.J. Jin and Prof. E-J Guo
Department of Physics & Center of Materials Science and Optoelectronics Engineering, University of Chinese Academy of Sciences, Beijing 100049, China

T.L. Deng and Prof. C. Liu
Beijing Synchrotron Radiation Facility, Institute of High Energy Physics, Chinese Academy of Sciences, Beijing 100049, China

Prof. T. Zhu
Songshan Lake Materials Laboratory, Dongguan, Guangdong 523808, China

Prof. L. Gu
National Center for Electron Microscopy in Beijing and School of Materials Science and Engineering, Tsinghua University, Beijing 100084, China

Prof. J. Chen
Hainan University, Haikou, 570228, Hainan Province, China

**Abstract:**

Creating a heterostructure by combining two magnetically and structurally distinct ruthenium oxides is a crucial approach for investigating their emergent magnetic states and interactions. Previously, research has predominantly concentrated on the intrinsic properties of the ferromagnet $SrRuO_3$ and recently discovered altermagnet $RuO_2$ solely. Here, we engineered an ultrasharp sublattice-matched heterointerface using pseudo-cubic $SrRuO_3$ and rutile $RuO_2$, conducting an in-depth analysis of their spin interactions. Structurally, to accommodate the lattice symmetry mismatch, the inverted $RuO_2$ layer undergoes an in-plane rotation of 18 degrees during epitaxial growth on $SrRuO_3$ layer, resulting in an interesting and rotational interface with perfect crystallinity and negligible chemical intermixing. Performance-wise, the interfacial layer of 6 nm in $RuO_2$ adjacent to $SrRuO_3$ exhibits a nonzero magnetic moment, contributing to an enhanced anomalous Hall effect (AHE) at low temperatures. Furthermore, our observations indicate that, in contrast to $SrRuO_3$ single layers, the AHE of $[(RuO_2)_{15}/(SrRuO_3)_n]$ heterostructures shows nonlinear behavior and reaches its maximum when the $SrRuO_3$ thickness reaches tens of nm. These results suggest that the interfacial magnetic interaction surpasses that of all-perovskite oxides (~5-unit cells). This study underscores the significance and potential applications of magnetic interactions based on the crystallographic asymmetric interfaces in the design of spintronic devices.

**Main text**

1. **Introduction**

Interfacial interactions are central to the operation of virtually all fuctional heterostructures.[1-6] Magnetic heterostructures have attracted intense attention due to the emergent magneto-transport properties and exotic spin configurations stemming from the reconstruction of chemical composition, lattices, electrons and orbitals near the interface.[7-11] Over the past decades, perovskite-type transitional metal oxides (TMO) have been favored for constructing all kinds of oxide heterostructures,[3, 11, 12] where the high-quality interfaces are generally achieved between isostructural materials with identical crystallographic symmetry and parameters.[13-16] The critical effective thickness in such heterostructure is typically concentrated within 5-unit cells near the interface, leading to an overdependence of limited properties on the heterogeneous interface.[17-20] Constructing sublattice-matched heterointerface between two functionally and structurally distinct materials should be a focus direction in the future studies.

Rutile-structured $RuO_2$ (RO) has long been considered a Pauli paramagnetic semimetal, with prior research focusing on band structure descriptions of its transport properties, optical, and photoemission spectra.[21-23] Only afterwards has it been confirmed to be an itinerant antiferromagnetic (AFM) ground state with collinear AFM ordering.[24, 25] Recently, RO is more popular due to the anomalous Hall effect under large magnetic fields and specific orientations, which breaks the intuition that AHE is an intrinsic magnetoelectronic response of ferromagnets alone. Unlike conventional collinear antiferromagnets such as $Mn_2Au$ with highly symmetric magnetization isosurfaces, spin-split energy bands, broken time-reversal and space-inversion symmetry of RO provide a new dimension to induce rich physical phenomenon.[26-28] Owing to the high dependence of spin-electron configuration on the distribution of nonmagnetic atoms, reaearchers have modified its magneto-electric effect through chemical doping and construction of heterostructures with Co, Py, etc. [26, 27, 29-32] However, numerous defects and irregular matching mode at amorphous interfaces are detrimental to investigate spin-orbit coupling, exchange interactions, and spin-polarization effects. It is imperative for us to search an ingenious and purer method that significantly change the crystal structure and coordinate environment in RO.

In this work, we chose the perovskite-type $SrRuO_3$ (SRO) provided tensile strain and rotational effect, paired with rutile-structure RO to construct ferromagnetic/antiferromagnetic all-ruthenium oxide heterostructures. Overcoming the unstable heterointerface with large structural and/or chemical strain induced by different crystallographic structures, RO/SRO heterostucture revealed an 18 degree in-plane rotation of RO to form a smooth and highly epitaxial interface. In view of the anion positions and hybridization strength between Ru and O in RO may alter significantly, the interfacial RO layer of 6 nm adjacent to SRO shows a nonzero magnetic moment, contributing to an enhanced AHE signal at low temperatures. Magnetic interaction in the SRO/RO bilayer results in a maximum AHE when the SRO thickness reaches 11 nm, the result is much greater than the depth of interfacial interactions of 5-unit cells in conventional perovskites oxide heterostructures.[15, 20, 33-37] This highlights the importance of magnetic interactions based on asymmetric interfaces in the spintronics design.

## 2. Results and Discussion

### 2.1 Structural characterizations of the $RuO_2/SrRuO_3$ heterostructure

The SRO has a pseudo-cubic structure with $a_{pc}$= 3.93 Å, while rutile-type RO shows a tetragonal structure with $a = b = 4.49$ Å, $c = 3.10$ Å). To grow the above materials epitaxially on $SrTiO_3$ (STO) substrate simultaneously, a suitable matching mode and the smallest interfacial strain are required due to the large difference in group symmetry and lattice parameters. We noticed that when the $c$-axis of RO is inverted into $ab$ plane and matched in the film plane as (101) RO // (110) SRO // (110) STO, the in-plane mismatch strain between the two materials reduces from 12% to 2.5%, as shown in Figure 1a and 1b. This provides crystallographic support for the construction of the heterogeneous interfaces between those ruthenium oxides. Using pulsed laser deposition technique (PLD), we fabricated a RO/SRO heterostructure on STO substrates. Figure 1c displays the XRD θ-2θ scans of RO/SRO heterostructure, the sharp diffraction peaks and clear Laue oscillations observed around the film's peaks are indicative of the high crystalline quality and epitaxial growth of all the films. The out-of-plane lattice parameters of SRO and RO are 3.953 ± 0.005 Å and 4.480 ± 0.005 Å, subjected to misfit strain of 0.6% and −0.2%, respectively. To demonstrate the epitaxial relationship between the RO/SRO heterostructure and substrate, we performed phi-scans around the (101) planes of STO, SRO, and the inverted RO, respectively. As shown in Figure 1e, SRO and STO have the same fourfold symmetry, whereas the in-plane symmetry of inverted RO shows a significantly difference (Figure S1, Supporting Information). According to the principle of in-plane matching of (101) RO // (110) SRO, we plotted the possible matching mode in Figure 1d. Twelve peaks in RO's (101) phi scan represent three sets of fourfold-symmetric periods, which corresponds to the red, blue, and green colored cells in RO. In this case, a twinned crystal is formed between red cells and blue cells at an angle of 24.9°, corresponding to the two peaks with a spacing of ~ 26.5° with the same intensity in the phi scan. Additionally, the region formed by the diagonals of the red and blue cells leads to the third symmetric period (green cells). The angle from the edge of blue cells to the diagonal of red cells is 32.55°, corresponding to the two peaks with a spacing of ~31.5° with different intensity in the phi scan. Since the rotation angles in Figure 1d are all calculated by undistorted unit cells, there are small deviations from the strained films. These characterizations unfold an interesting and rotational stacking mode between the oxide heterointerface, which may add a modulated dimension to the exploration for physical properties.

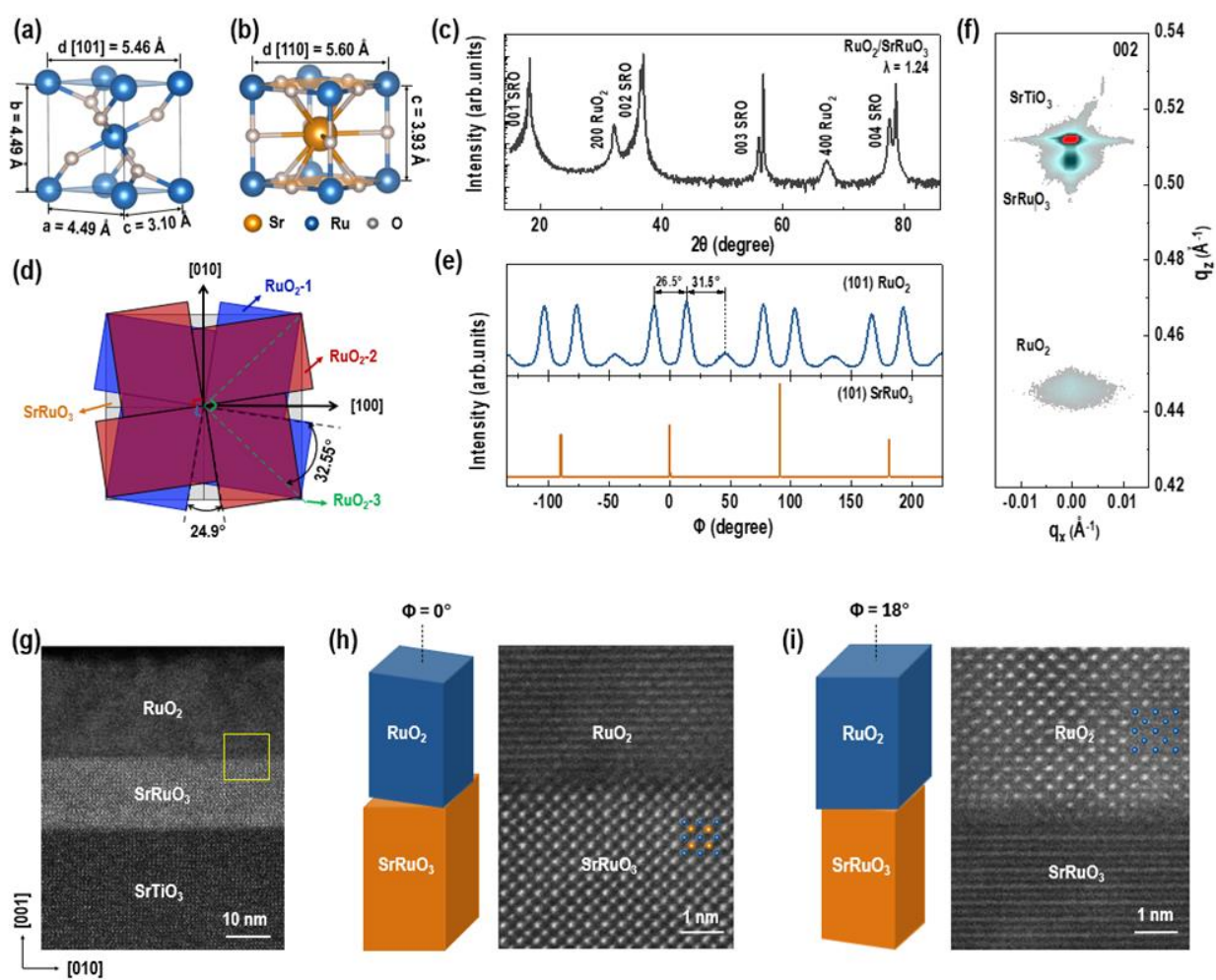

**Figure 1. Structural characterizations of the RuO$_2$/SrRuO$_3$ heterostructure.** Schematics of (a) SrRuO$_3$ pseudo-cubic unit cell and (b) RuO$_2$ tetragonal unit cell. (c) The XRD $\theta$-$2\theta$ scan of the RuO$_2$/SrRuO$_3$ heterostructure. (d) In-plane lattice matching mode along [110] and [101] directions of SrRuO$_3$ and RuO$_2$, respectively. Grey squares represent SrRuO$_3$ crystal cells, blue, red and green rectangles represent RuO$_2$ crystal cells, each color corresponding to three different symmetric periods. (e) The Phi scans around (101) planes of RuO$_2$ and SrRuO$_3$ layers. (f) RSM around 002 diffraction peak of SrTiO$_3$ substrates. (g) Low-magnified cross-sectional STEM image of an RuO$_2$/SrRuO$_3$ heterostructure grown on SrTiO$_3$ substrates. (h, i) High-resolution HADDF-STEM image near the RuO$_2$/SrRuO$_3$ interface viewed along the [100] zone axis

of SrRuO$_3$ and RuO$_2$ layer. The left side of each panel shows the corresponding structural model, demonstrating the observation direction is rotated 18° along the *c*-axis.

Figure 1f presents reciprocal space mapping (RSM) near the (002) plane of the STO substrate, the in-plane strain of heterostructure is fully constrained by STO substrate whereas out-of-plane is consistent with the results of θ-2θ scan. We additionally conducted cross-sectional high-angle annular dark field scanning transmission electron microscopy (HAADF-STEM) images along the [100] zone axis, as depicted in Figure 1g. It can be noted that clear interfaces and atomic arrangements between the RO/SRO heterostructure and STO substrate demonstrate high crystallinity and perfect epitaxial relationship of all film. In order to visualize atomic-resolved STEM images of SRO layer (Figure 1h) and RO layer (Figure 1i), we have to rotate the specimen by 18° around axis and take the measurement separately. This rotation angle is accordant with the results of the phi scan. In the SEAD pattern of RO/SRO heterostructure, the crystal planes represented by diffraction spots are labeled on the pattern, the strain states along the out-of-plane calculated by plane spacing are consistent with the results of XRD θ-2θ scan and RSM mapping (Figure S2, Supporting Information). The structural analysis demonstrates that the heterogeneous interfaces between RO and SRO are atomically sharp, chemically uniform, and with high crystallographic quality.

**2.2 Exchange bias in the RuO$_2$/SrRuO$_3$ heterostructure**

Next, we examine the exchange bias (EB) effect of RO/SRO heterostructure since it is crucial to understand the magnetic exchange coupling at the interfaces.[38-42] Field-dependent magnetization (*M-H*) measurements were performed at various temperatures. Figure 2a shows the *M-H* loops of the RO/SRO heterostructure under different cooling processes. Unlike zero-field cooling hysteresis loop (gray area), the *M-H* loops are apparently not symmetric after cooling under positive (red curve) and negative (blue curve) magnetic fields. Considering the presence of EB effect near the antiferromagnetic/ferromagnetic interface, we extracted the zero-field-cooling (ZFC) *M-H* curve from the field-cooling (FC) *M-H* curves. Clearly, the RO/SRO heterostructure exhibits the pinning effect of ferromagnetic SRO by the antiferromagnetic RO after a change in the direction of magnetic fields. Figure S3c shows the *M-H* measurements at various temperatures, all curves show the negative EB effect similar to previous studies, which disappears when the temperature above Curie temperature ($T_C$) of SRO (~150 K).[44] And then we summarize the temperature dependency of the saturated fields ($H_S$) and coercive fields ($H_C$) in Figures 2c and 2d, respectively, both $H_S$ and $H_C$ decrease as increasing temperature. Furthermore, the *M-H* loops display an open configuration, with their centers displaced to negative (or positive) fields by the exchange bias field ($H_{EB}$) of approximately 224 Oe at 10 K (Figure S3, Supporting Information). The temperature dependencies under both positive and negative fields are summarized in Figure 2e. Above results indicate that these interfacial pinned moments are strongly coupled to $Ru^{4+}$ moments of the antiferromagnetic spin state, leading to a significant EB effect. The negative $H_{EB}$ implies a ferromagnetic double exchange interaction $Ru^{4+}$-$O^{2-}$-$Ru^{4+}$ between antiferromagnetic (AFM) and ferromagnetic (FM) layers.[11, 43-45]

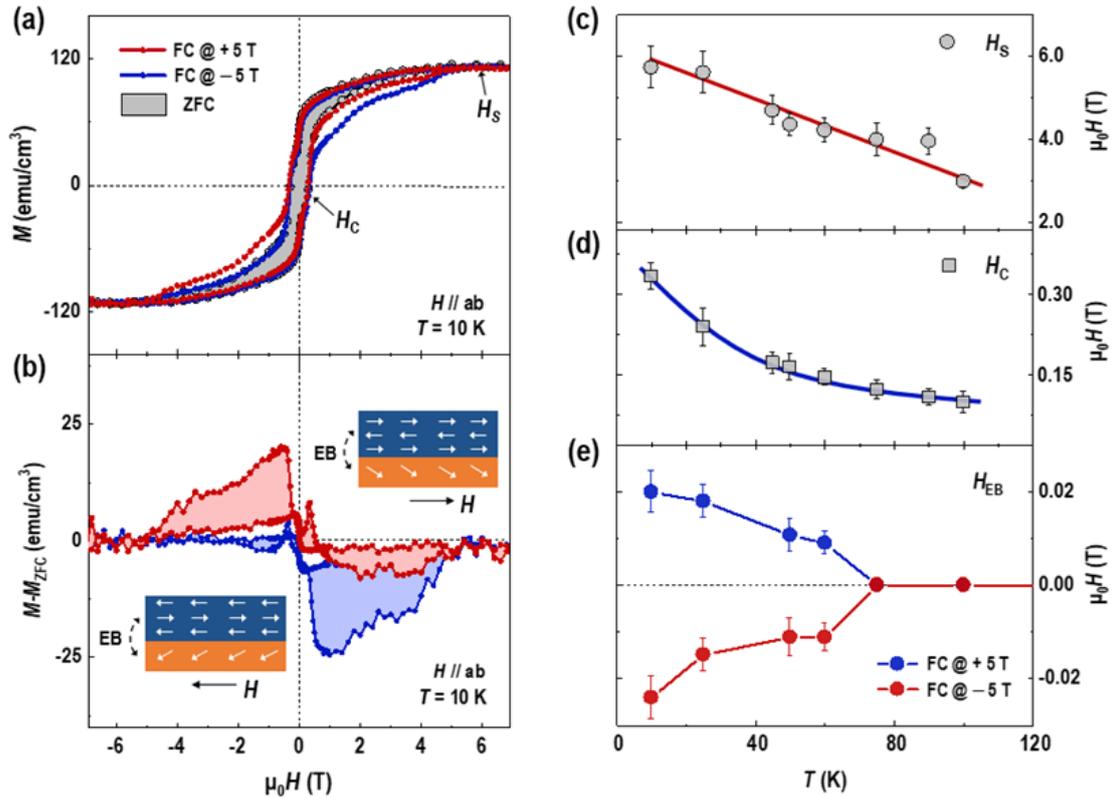

**Figure 2. Exchange bias observed in the RuO$_2$/SrRuO$_3$ heterostructure.** (a) *M-H* hysteresis loops at 10 K under in-plane magnetic field. The red and blue curves represent *M-H* hysteresis loops measured after field-cooling in +5 T and −5 T, respectively. The shadow region shows *M-H* hysteresis loop measured after zero field cooling. (b) The subtracted $M-M_{ZFC}$ as a function of the field at 10 K under in-plane magnetic field. The insets show schematics of the spin alignment across the heterogeneous interfaces, demonstrating the interface exchange bias effect between RuO$_2$ and SrRuO$_3$ layers. (c-e) The temperature (T) dependence of saturated field ($H_S$), coercive field ($H_C$), and exchange bias field ($H_{EB}$) is derived from the analysis of *M-H* hysteresis loops at different temperatures.

## 2.3 Magnetic and structural profile in the RuO$_2$/SrRuO$_3$ heterostructure

To explore the modulated effect of interface on the magnetic behaviours of the RO layer, we conducted polarized neutron reflectivity (PNR) measurements to identify magnetic depth profile across the entire film. The specular neutron reflectivities, expressed as a function of the wave factor q = 4πsinθ/λ, were measured for both spin-up ($R^+$) and spin-down ($R^−$) polarized neutrons. These reflectivities were normalized to the asymptotic value of the Fresnel reflectivity ($R_F = 16\pi^2/q^4$), where θ is the incident angle, and λ is the wavelength of incident neutrons (Figure S4, Supporting Information). Figure 3a depicts the calculated spin asymmetry along with its corresponding fit, which allows for the derivation of both chemical and magnetic depth profiles (Figure 3b). The magnetic depth profile reveals that the SRO layer maintains a stable magnetization of approximately 40 emu/cm$^3$, with a slight decrease observed near the interface with the RO layer. Obviously, it can be found that the RO layer is separated into two parts: top and bottom layers, which are determined by distance to the RO/SRO interface. The RO layer near the interface has non-zero magnetic moments of roughly 15 emu/cm$^3$. Combined with the results of nSLD, the thickness of RO with non-zero magnetic moments is nearly 6 nm (~13 unit cells), which far exceeds the magnetic depth influenced by the mostly interfaces between conventional isostructural oxides (~5 unit cells), such as heterostructure between SrRuO$_3$ with SrTiO$_3$, SrCuO$_3$, SrMnO$_3$ and PrMnO$_3$ .[18-20, 46-48] The field-cooled PNR data suggests that considerable thickness of RO is magnetized due to the exchange coupling and ferromagnetic pinning effect near the heterogeneous interface.

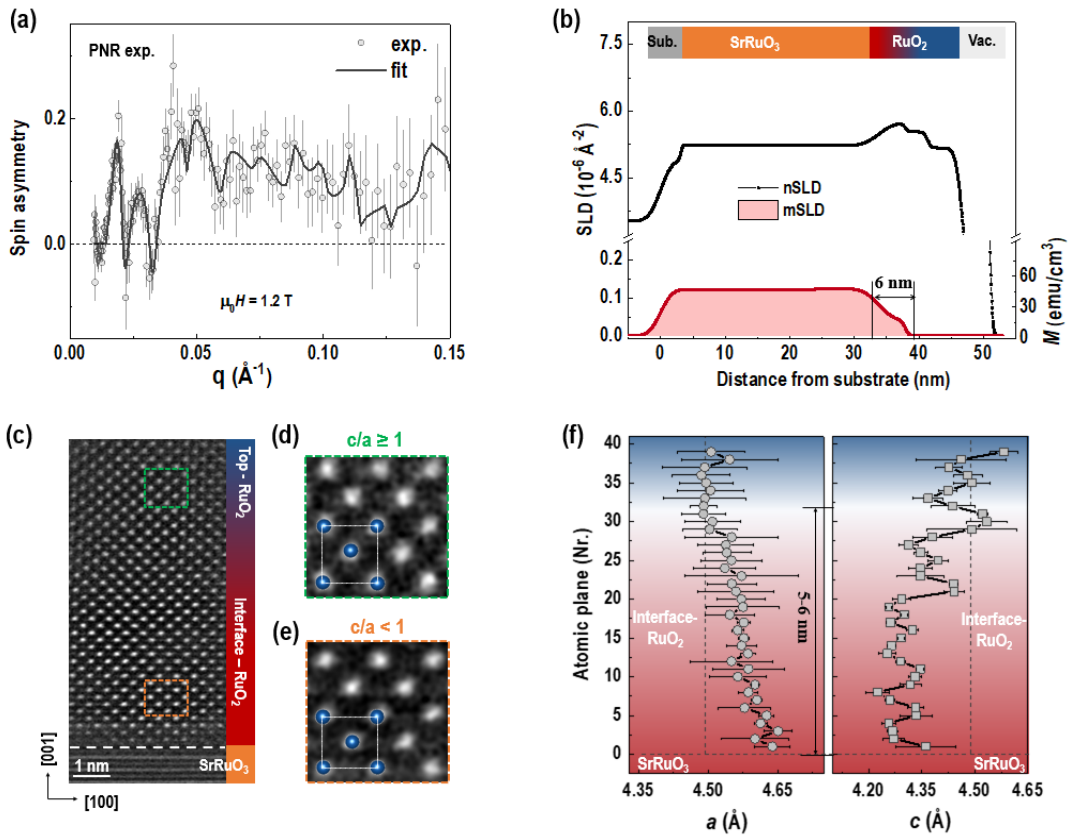

**Figure 3. Considerable magnetic depth in RuO$_2$ layer near the interface.** (a) Calculated neutron spin asymmetry (SA) as $(R^+ - R^-)/(R^+ + R^-)$, where $R^+$ and $R^-$ are the normalized neutron reflectivities from spin-up and spin-down polarized neutrons, respectively. The open symbols denote the experimental data, while the solid lines correspond to the best-fit curves. (b) The black line and red line represent nuclear scattering

length density (*n*SLD) and magnetic scattering length density (*m*SLD) depth profiles of the $RuO_2$/$SrRuO_3$ heterostructure, respectively. (c) High-resolution HADDF-STEM image near the $RuO_2$/$SrRuO_3$ interface viewed along [100] zone axis of the $RuO_2$ layer. The magnified representative unit cells (e) near and (d) away from the heterogeneous interface, marked in green and orange dashed squares, respectively. (f) The layer position dependence of in-plane and out-of-plane lattice parameters of the $RuO_2$ layer, with the horizontal and vertical dashed lines representing the $RuO_2$/$SrRuO_3$ interface and lattice parameters of strain-released $RuO_2$, respectively.

As mentioned above, we analyzed microstructural information in RO to demonstrate the origin of unusual magnetic behavior. Figure 3c shows high-resolution HADDF-STEM image near the RO/SRO interface viewed along [100] zone axis of RO layer. The representative STEM images from the top and bottom in RO layer are shown in Figures 3d and 3e, respectively. The schematics intuitively illustrate that the unit cells of bottom RO layer are elongated laterally with $c/a < 1$, whereas that of top RO layer are elongated vertically with $c/a \geq 1$. As is shown in Figure 3f, we further summarized the layer position-dependent lattice parameters of RO layers, which were calculated by average spacing of the atom columns in all atomic plane. It can be seen that the lattice distortion in RO is not released until ~12 unit cells (~5-6 nm) away from the interface. This result is consistent with the thickness of ferromagnetic RO layer obtained from the PNR measurements. We believe that the combination of interfacial strain and hetero-structural stacking mode plays a crucial role in considerable depth effected by the sublattice-matched heterointerface.

### 2.4 XAS characterizations of the $RuO_2$/$SrRuO_3$ heterostructure

The electronic states are strongly correlated with their magnetic behaviors, we performed XAS measurement for antiferromagnetic RO film grown on (100)-oriented $TiO_2$ substrate with identical structure and parameters.[49] As a comparison, we also fabricated [$(RuO_2)_m$/$(SrRuO_3)_{30}$] ($R_mS_{30}$) heterostructures with different thicknesses of RO layer ($m$ = 6, 40 nm) (Figure S5, supporting information). The electronic states of RO in $R_6S_{30}$ heterostructure are affected by rotational stacking mode and interfacial strain, whereas that in $R_{40}S_{30}$ heterostructure is solely affected by rotational stacking mode. In Figure 4a, the intensity of two peaks in XAS data represents the degree of hybridization between O-2$p$ and Ru-4$d$ $t_{2g}$/$e_g$ orbitals, respectively. The higher intensity indicates fewer electron-occupied states in the Ru-4$d$ orbital, and then we defined the difference between the first and second peaks as $I(t_{2g}-e_g)$. Figure 4b shows that I(RO) < I($R_{40}S_{30}$) < I($R_6S_{30}$), illustrating an increasement of the electron-occupied states on the Ru-4$d$ $e_g$ orbital. Furthermore, the electrons will be excited more easily to the $e_g$ orbitals in $R_6S_{30}$ heterostructure due to the crystal field splitting energy is significantly smaller than that of $R_{40}S_{30}$ heterostructure ($\Delta C_{F1} < \Delta C_F$). Above analyses give solid evidence for the ferromagnetic layer of 6 nm in interfacial RO from the view of electronic occupation. With the results of XAS, the energy level diagrams of Ru 4$d$ in bulk RO and R$_m$S$_{30}$ samples are shown in Figures 4c and 4d, respectively. It is highly possible that the partial electrons may jump from Ru-4$d$ $t_{2g}$ orbitals to $e_g$ orbitals, breaking the original long-range AFM ordering and leading to the FM ordering in the interfacial RO layer. These results consistently agree with the studies reasearched by Herklotz[50] and Jeong[51] et al., demonstrating that the spin state and electric structure of Ru atom are effectively modulated by crystallographic parameters associated with the adjacent oxygen coordination.

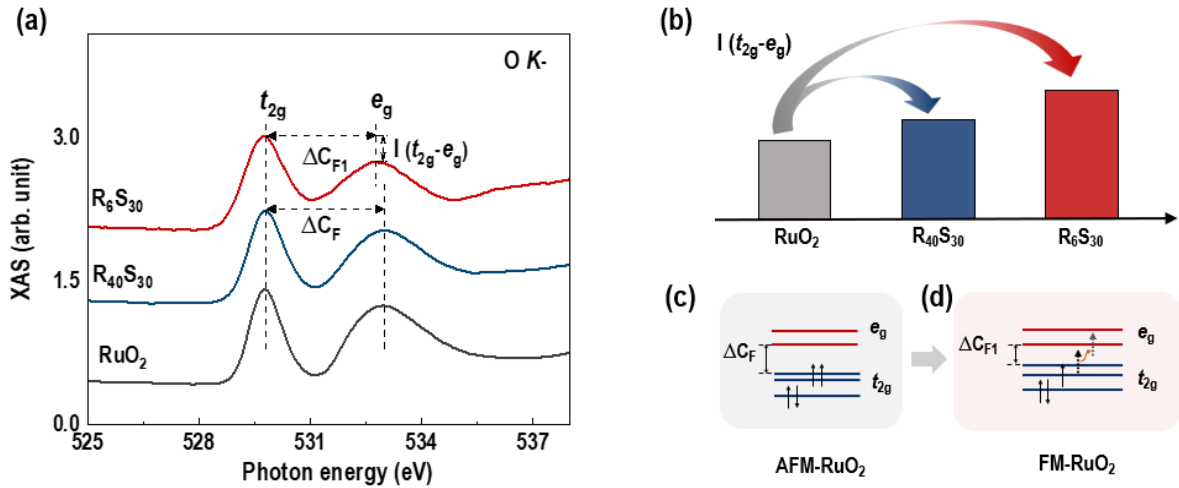

**Figure 4. XAS characterizations of the RuO$_2$/SrRuO$_3$ heterostructure.** (a) XAS of O $K$-edges for R$_6$S$_{30}$, R$_{40}$S$_{30}$ heterostructures on SrTiO$_3$ substrate and RuO$_2$ single layer film on TiO$_2$ substrate with identical orientation, respectively. (b) Difference in the hybridization between O-2$p$ and Ru-4$d$ $t_{2g}$/$e_g$ for R$_6$S$_{30}$, R$_{40}$S$_{30}$ heterostructures and RuO$_2$ single layer film, derived from the intensity of the first and second peaks in XAS data. (c, d) Single particle energy level diagrams of the Ru-4$d$ orbitals for bulk RuO$_2$ and the RuO$_2$ regulated by our heterogeneous interface.

## 2.5 Enhanced AHE in the RuO$_2$/SrRuO$_3$ heterostructure

AHE is a typical property of ferromagnets, it is widely believed believed that both the intrinsic mechanisms associated with Berry curvature and extrinsic spin-dependent scattering mechanisms can contribute to AHE observed in SRO.[52-54] Given the changes in magnetic spin states and electronic structure of RO layer, we measured the Hall resistivity ($\rho_{xy}$) of heterogenous films using a conventional setup (Figure 5a). Typically, the Hall resistivity ($\rho_{xy}$) in a ferromagnet can be caculated by $\rho_{xy} = R_0H + R_SM_z$, where R$_0$ is the ordinary Hall effect coefficient, R$_S$ is the anomalous Hall effect coefficient, and $M_z$ is the magnetization.[55] We isolated the anomalous Hall effect value from the total Hall resistivity ($\rho_{xy}$) by subtracting the ordinary Hall resistivity (R$_0$H). Figure 5b displays the field-dependent ($\rho_{xy} - $R$_0$H) (($\rho_{xy} - $R$_0$H)-$H$) for RO/SRO heterostructure and SRO sigle film at 10 K. It is evident that the saturation value of ($\rho_{xy} - R_0H$) in the heterostructure is 0.24, enhancing by 74% compared to that of SRO single film with the same thickness of SRO. The ($\rho_{xy} - R_0H$)-$H$ curves at various temperatures exhibit a similar difference (Figure 5e), displaying a parabolic shape with increasing temperature and disappearing when the temperature reaches 150 K, as indicated by the gray line in Figure 5c. Recent studies have revealed that RO is a collinear antiferromagnet, characterized by low symmetric magnetization isosurfaces that are modulated by the hybridization of antiferromagnetic Ru and nonmagnetic O orbitals within the oxygen octahedron. Thus, the anomalous Hall effect not only originates in conventional ferromagnets but may also arise from low symmetric crystal structures and distinct microscopic origins in collinear antiferromagnets. The Hall effect's dependence on the distribution of nonmagnetic atoms offers a strategic approach in the design of heterostructures with strong magnetic interactions. In our work, the different crystal symmetries between SRO and RO, strong exchange coupling effects, and mutual strain near the heterogeneous interface all exacerbate the breaking of inversion symmetry in RO and the hybridization between Ru-O, resulting in an enhanced anomalous Hall effect (AHE) at low temperatures.[9, 54, 56]

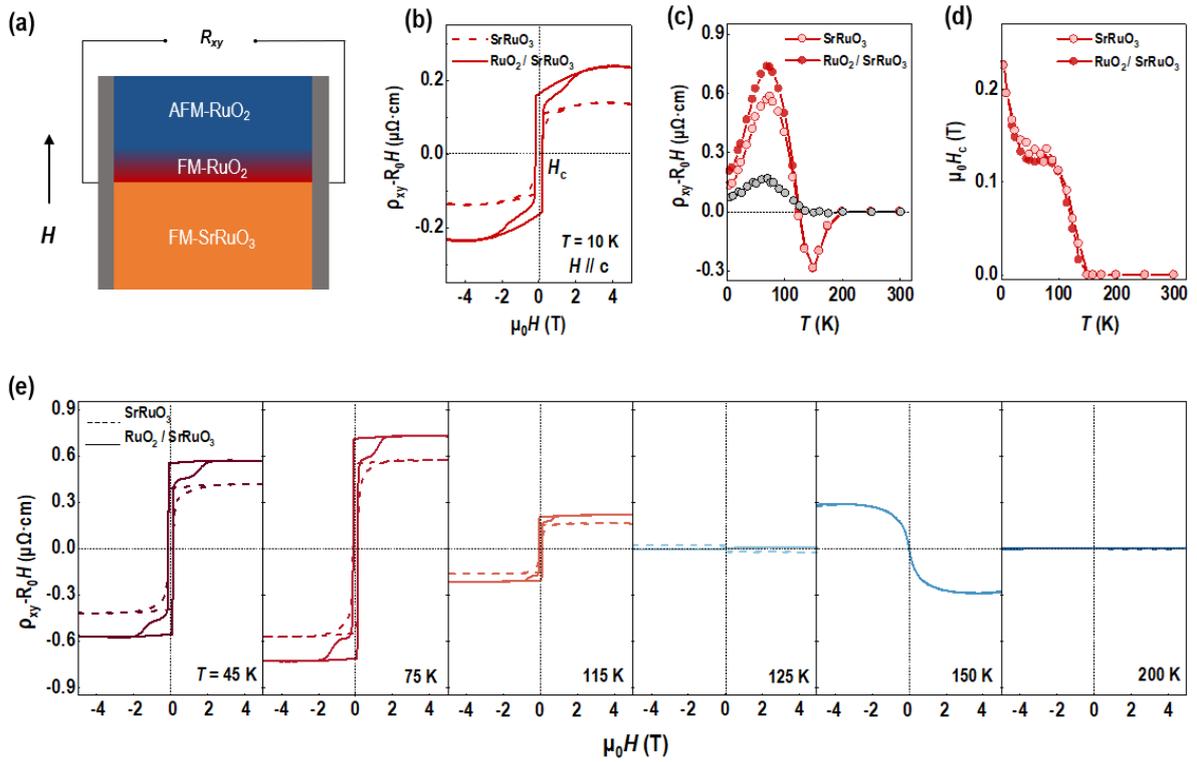

**Figure 5. Enhanced AHE in the RuO₂/SrRuO₃ heterostructure.** (a) Schematic of Hall measurements on the RuO₂/SrRuO₃ heterostructure. (b) Enhanced anomalous Hall resistivity ($\rho_{xy} - R_0H$) of the RuO₂/SrRuO₃ heterostructures compared to that of the SrRuO₃ single layer film. (c) The $\rho_{xy} - R_0H$ and (d) coercive fields ($H_C$) as a function of temperature. The grey dots are the difference in $\rho_{xy} - R_0H$ between the SrRuO₃ single layer films and the RuO₂/SrRuO₃ heterostructures.

## 2.6 Critical thickness of RuO$_2$ and SrRuO$_3$ driven by the heterointerface

To delve deeper into understanding the critical effective thickness of RO/SRO interface and its impact on the physical properties of materials on both sides, we fabricated two sets of heterostructures [(RuO$_2$)$_{15}$/(SrRuO$_3$)$_n$] (R$_{15}$S$_n$) and [(RuO$_2$)$_m$/(SrRuO$_3$)$_{30}$] (R$_m$S$_{30}$) with different thicknesses of SRO (*n*) and RO (*m*) (Figures S5 and S6, Supporting Information). Concurrently, we prepared several SrRuO$_3$ (S$_n$) single layer films with matched SRO thicknesses for comparison. Magneto-transport measurements were conducted on both R$_{15}$S$_n$, R$_m$S$_{30}$ heterostructures, and S$_n$ single layer films. Figures 6a and S7 shows the $\rho-T$ curves of R$_{15}$S$_n$ heterostructures and S$_n$ single layer films. In contrast to the Metal-Insulator Transformation (MIT) observed in the 2-nm-thick SRO single layer film, all heterostructures exhibit metallic behavior across temperatures. Moreover, the resistivity of all S$_n$ single layer films is nearly four times larger than that of R$_{15}$S$_n$ heterostructures. Figure 6b reveals a non-monotonic variation in the thickness-dependent $\rho$ of SRO layer. Notably, there is a peak at R$_{15}$S$_{11}$ heterostructure, indicating significant modulation of the electrical transport properties near the interface due to carrier migration between SRO and RO.[48, 57]

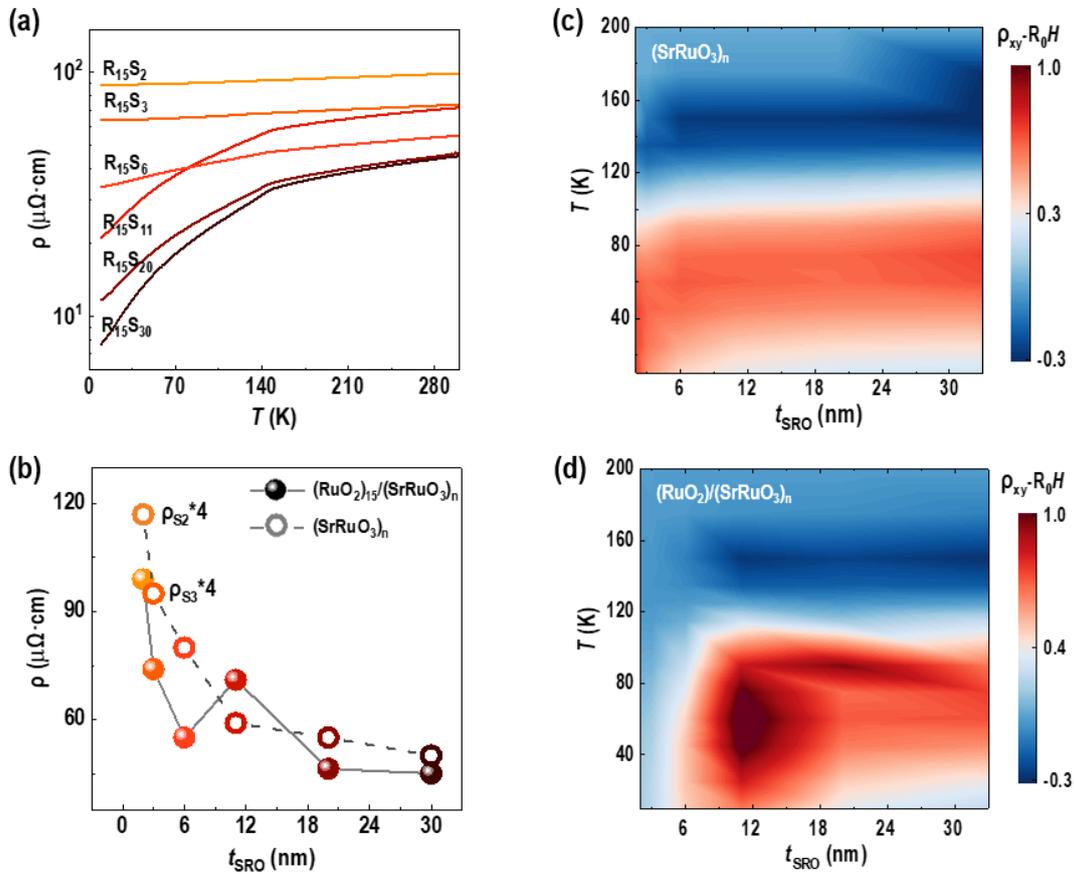

**Figure 6. Nonlinear AHE cofficient and critical thickness in [(RuO$_2$)$_{15}$/(SrRuO$_3$)$_n$] (R$_{15}$S$_n$) heterostructures.** (a) $\rho-T$ curves of R$_{15}$S$_n$ heterostructures. (b) The room-temperature resistivity as a function

of SrRuO$_3$ layer thickness. The $\rho_{xy}$–R$_0$H as a function of temperature and SrRuO$_3$ layer's thickness ($t_{SRO}$) for (c) the SrRuO$_3$ single layer films and (d) the R$_{15}$S$_n$ heterostructures.

Furthermore, we conducted measurements of ($\rho_{xy}$ − R$_0$H) as a function of fields for all R$_{15}$S$_n$ heterostructures and S$_n$ single layer films at various temperature (Figure S8, Supporting Information). Subsequently, we quantified all ($\rho_{xy}$ − R$_0$H)-H curves as three-dimensional phase diagrams of ($\rho_{xy}$ − R$_0$H) dependent temperature and SRO layer's thickness for R$_{15}$S$_n$ heterostructures (Figure 6d) and S$_n$ single layer films (Figure 6c), respectively. The phase diagrams reveal that HC reduced toward zero with increasing temperature and thickness of SRO for all R$_{15}$S$_n$ bilayers and S$_n$ single layer films. It is evident that the difference of ($\rho_{xy}$ − R$_0$H) between heterostructures and S$_n$ single layer films is indicated by the color contrast. In both S$_n$ single layer films and heterostructures, there is a consistent trend for ($\rho_{xy}$ − R$_0$H) to increase and then decrease with raising temperature, accompanied by a switchable sign of ($\rho_{xy}$ − R$_0$H). This phenomenon aligns with previous studies,[7, 10, 16] suggesting a correlation to the sensitive energy dependence of ($\rho_{xy}$ − R$_0$H), where changes in momentum space near the Fermi level easily affect both the magnitude and sign of ($\rho_{xy}$ − R$_0$H).[58] Additionally, at the temperature of 75 K, ($\rho_{xy}$ − R$_0$H) in heterostructures exhibits a nonlinear variation with increasing thickness compared to S$_n$ single layer films, showing a parabolic trend and reaching maximum at the SRO layer's thickness of 11 nm (Figure S9a, supporting information). Furthermore, the electrical transport curves, three-dimensional phase diagrams of ($\rho_{xy}$ − R$_0$H) dependent on temperature and RO layer's thickness for R$_m$S$_{30}$ heterostructures is similar to that of R$_{15}$S$_n$ (Figure S10, supporting information). Moreover, as the thickness of RO increases, ($\rho_{xy}$ − R$_0$H) also exhibits a parabolic trend, reaching a maximum at the RO thickness of 6 nm.

These results demonstrate that AHE of the R$_m$S$_{30}$ bilayer reaches its peak at a moderate thickness of the RO layer (~15 nm), with excessive thickness leading to a weakening of the total Hall signal. The regulation of the RO layer near the interfacial region results in a combined effect that decreases the overall ($\rho_{xy}$ − R$_0$H) value with increasing RO thickness. Similarly, the AHE of the R$_{15}$S$_n$ bilayer reaches maximum value at a moderate thickness of the SRO layer (~11 nm), with both excessively thinner and thicker layers diminishing the total Hall effect. This observation suggests that the influence of the RO layer primarily affects the nearby depths of the SRO layer, particularly in its most disordered state. At an optimal thickness, the SRO layer withstands the strain from both the substrate and the top RO layer, leading to significant scattering and the highest anomalous Hall effect. Conversely, excessively thick SRO layers exhibit a large fraction of bulk state, weakening the anomalous Hall resistivity. And excessively thin SRO layers result in reduced Hall resistivity due to complex strain states and shorter scattering paths. Consequently, the heterogeneous interfaces not only modulate the RO layer but also the mutual strain and exchange coupling interactions affecting the magnetoelectric properties of the SRO layer. The depth of magnetic interaction at the SRO/RO interface surpasses that of conventional isostructural materials.

## 3. Conclusion

In conclusion, we fabricated the RO/SRO heterostructures, the RO layer undergoes an in-plane rotation of 18 degrees during the epitaxial growth on SRO, resulting in an atomically smooth and highly crystalline interface with negligible chemical intermixing. Due to the lattice structure and degree of hybridization between Ru and O altering significantly, the interfacial RO layer of nearly 6 nm adjacent to SRO exhibits non-zero magnetic moments, contributing to an enhanced anomalous Hall effect (AHE) at low temperatures. Additionally, our observations indicate that, in contrast to $SrRuO_3$ single layers, the AHE of heterostructures displays a nonlinear behavior and reaches its maximum when the SRO thickness reaches to 11 nm, suggesting that the unprecedentedly effective depth not only in $RuO_2$ layer but also in $SrRuO_3$ layer. The result is much greater than the depth of interfacial interactions of 5-unit cells in conventional perovskites oxide heterostructures. This study underscores the significance and advantages of coherent interfaces and well-defined orientations when integrating dissimilar materials into heterostructure stacking. Textured growth of lattice-mismatched oxides offers new opportunities for stronger magnetic interactions to develop high-performance spintronic devices.

## 4. Experimental Section

*PLD synthesis of thin films:* The SRO single films and RO/SRO heterostructures of varying thicknesses on (001)-oriented STO substrates were fabricated by Pulsed Laser Deposition (PLD) technology. During the growth process, the laser fluence was meticulously maintained within the range of 1.5 to 2 J/cm². Concurrently, the substrate temperatures were carefully controlled at 700 °C for the SRO layers and 600 °C for the RO layers, respectively. The pressure in the deposition chamber needs to be reduced to ~1 x $10^{-7}$ Torr before the film can be warmed up. And the films were cooled to room temperature at -15 °C/min under 100 Torr after laser bombardment. Throughout the process, we employed X-ray reflectivity measurements to ascertain the growth thickness per laser pulse. By controlling the number of laser pulses, we successfully achieved the fabrication of heterogeneous films with varying thicknesses.

*Structural characterizations:* X-ray diffraction θ–2θ scans and reflectivity measurements were carried out using a Cu Kα1 radiation source on a Panalytical X'Pert3 MRD system. Phi scans and reciprocal space mappings were conducted at the BLU202 beamline of the Shanghai Synchrotron Radiation Facility. HAADF-STEM images of RO/SRO heterostructure were measured along the pseudo-cubic [100] zone axis using an aberration-corrected FEI Titan Themis G2 microscope. The specific interface strain and in/out-of plane lattice parameters were quantitatively determined by analyzing the atomic positions using Gaussian fitting.

*Magnetic and transport measurements:* The magnetic properties of samples were characterized by MPMS-3. *M-T* curves were obtained ranging from 10 to 300 K under the in-plane magnetic field of 1 T. The *M-H* loops were measured at $T$ =10, 25, 60, 100 and 150 K after subtracting the diamagnetic signals from the substrates. The exchange bias was recorded after the sample was field-cooled down to specific temperature under magnetic fields of ±7 T. The transport properties were measured using standard van der Pauw geometry with wire-bonding technique, and these measurements were carried out with the 9T Physical Property Measurement System (PPMS).

*PNR setup and measurements:* Polarized neutron reflectometry (PNR) experiments were conducted on the RO/SRO heterostructure at the Chinese Spallation Neutron Source. The sample was simultaneously measured at $T$=110 K under an in-plane magnetic field of 1.2 T. It can be determined that the nuclear and magnetic scattering length densities by calculating the spin asymmetry, defined as SA = ($R^+$- $R^-$) / ($R^+$ + $R^-$), and subsequently fitting the PNR data using the GenX software, where $R^+$ and $R^-$ represent spin-up and spin-down polarized neutrons, respectively.

*XAS measurements:* Elemental-specific X-ray Absorption Spectroscopy (XAS) measurements were carried out at O *K*-edges in all samples at the 4B9B beamline of Beijing Synchrotron Radiation Facility (BSRF). The sample's scattering plane is oriented at a 45° angle relative to the direction of the incident X-ray beam.

**Supporting information**

Supporting Information is available from the Wiley Online Library or from the author.

**Acknowledgements**

We thank the fruitful discussions with Prof. Cheng Song (Tsinghua University), Prof. Guoqiang Yu, Prof. Enke Liu (IOP-CAS) and Prof. Yuzhu Song. Dr. Tengyu Guo from Songshan Lake Mater. Lab provided help on magnetization measurements. This work was supported by the National Key Basic Research Program of China (Grant Nos. 2020YFA0309100), the National Natural Science Foundation of China (Grant No. 22235002, U22A20263, 52250308), the Beijing Natural Science Foundation (Grant No. JQ24002), the CAS Project for Young Scientists in Basic Research (Grant No. YSBR-084), the Guangdong-Hong Kong-Macao Joint Laboratory for Neutron Scattering Science and Technology (HT-CSNS-DGCD-0080/2021). PNR experiments were conducted at the Beamline MR of Chinese Spallation Neutron Source (CSNS), CAS.

**Conflict of Interest**

The authors declare no conflict of interest.

**Data Availability Statement**

The data that support the findings of this study are available from the corresponding author upon reasonable request.

# Supporting Information

**Title:** Confined Magnetization at the Sublattice-Matched Ruthenium Oxide Heterointerface

Yiyan Fan,† Qinghua Zhang,† Ting Lin, He Bai, Chuanrui Huo, Qiao Jin, Tielong Deng, Songhee Choi, Shengru Chen, Haitao Hong, Ting Cui, Qianying Wang, Dongke Rong, Chen Liu, Chen Ge, Tao Zhu, Lin Gu, Kuijuan Jin*, Jun Chen*, and Er-Jia Guo*

† Y. Y. Fan and Q. H. Zhang contributed equally to this work.
\* Correspondence and requests for materials should be addressed to J. C. and E. J. G. (emails: junchen@ustb.edu.cn and ejguo@iphy.ac.cn)

**This file includes:**

Figure S1. Structural characterizations of the $RuO_2/SrRuO_3$ heterostructure.

Figure S2. Microscopic view of the $RuO_2/SrRuO_3$ heterostructure.

Figure S3. Exchange bias effects observed at the $RuO_2/SrRuO_3$ interfaces.

Figure S4. X-ray reflectivity of the $RuO_2/SrRuO_3$ heterostructure.

Figure S5. Structural characterizations of $(RuO_2)_m/(SrRuO_3)_{30}$ ($R_mS_{30}$) heterostructures.

Figure S6. Structural characterizations of $(RuO_2)_{15}/(SrRuO_3)_n$ ($R_{15}S_n$) heterostructures.

Figure S7. Transport properties of $SrRuO_3$ ($S_n$) single films.

Figure S8. Anomalous Hall resistance of $SrRuO_3$ ($S_n$) single layer films and $(RuO_2)_{15}/(SrRuO_3)_n$ ($R_{15}S_n$) heterostructures at different temperatures.

Figure S9. Nonmonotonic AHE coefficient in the $RuO_2/SrRuO_3$ heterogeneous films with various thickness of $SrRuO_3$ layer.

Figure S10. Transport properties of $(RuO_2)_m/(SrRuO_3)_{30}$ ($R_mS_{30}$) heterostructures.

**Figures S1 to S11**

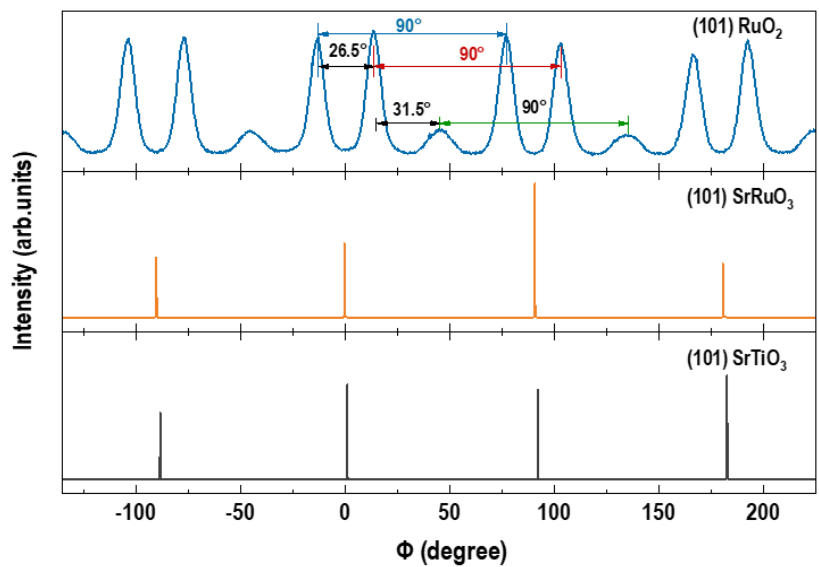

**Figure S1. Structural characterizations of the RuO₂/SrRuO₃ heterostructure.** Phi scans around (101) plane of $RuO_2$, $SrRuO_3$ and $SrTiO_3$.

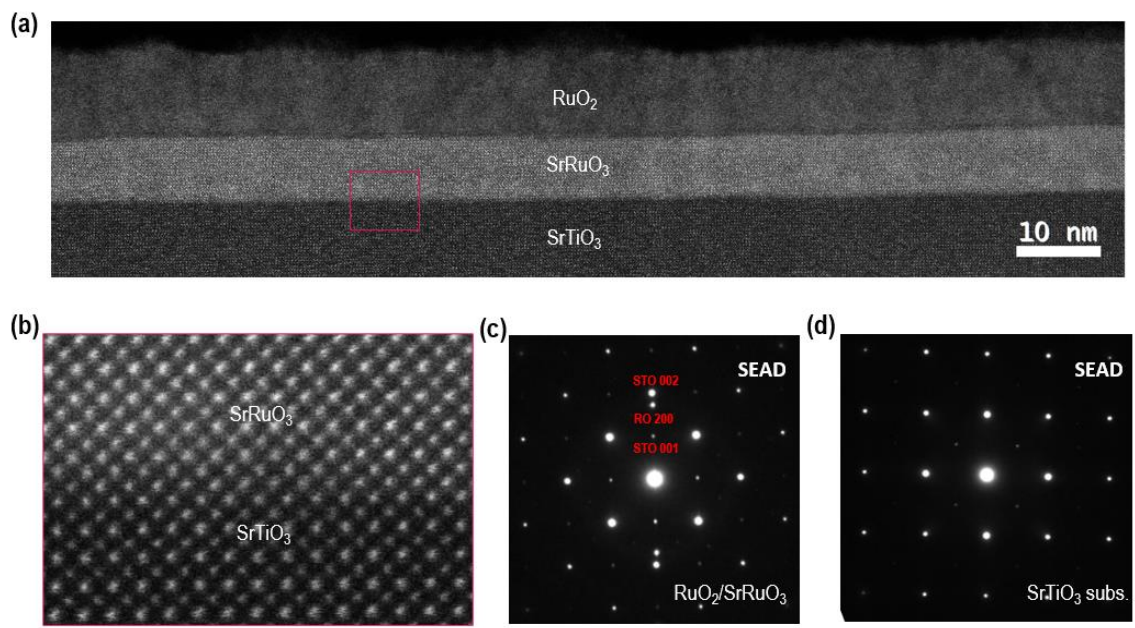

**Figure S2. Microscopic view of the RuO$_2$/SrRuO$_3$ heterostructure.** (a) Low-magnification STEM image of a RuO$_2$/SrRuO$_3$ heterostructure. The interface between RuO$_2$ and SrRuO$_3$ is continuous, and the surface is flat. (b) High-magnification STEM image of a representative region marked in (a). The atomic structure of SrRuO$_3$ is illustrated. The interfaces are atomically sharp with roughness below an atomic layer. (c) and (d) SEAD patterns of RuO$_2$/SrRuO$_3$ heterostructure and SrTiO$_3$ substrates, respectively.

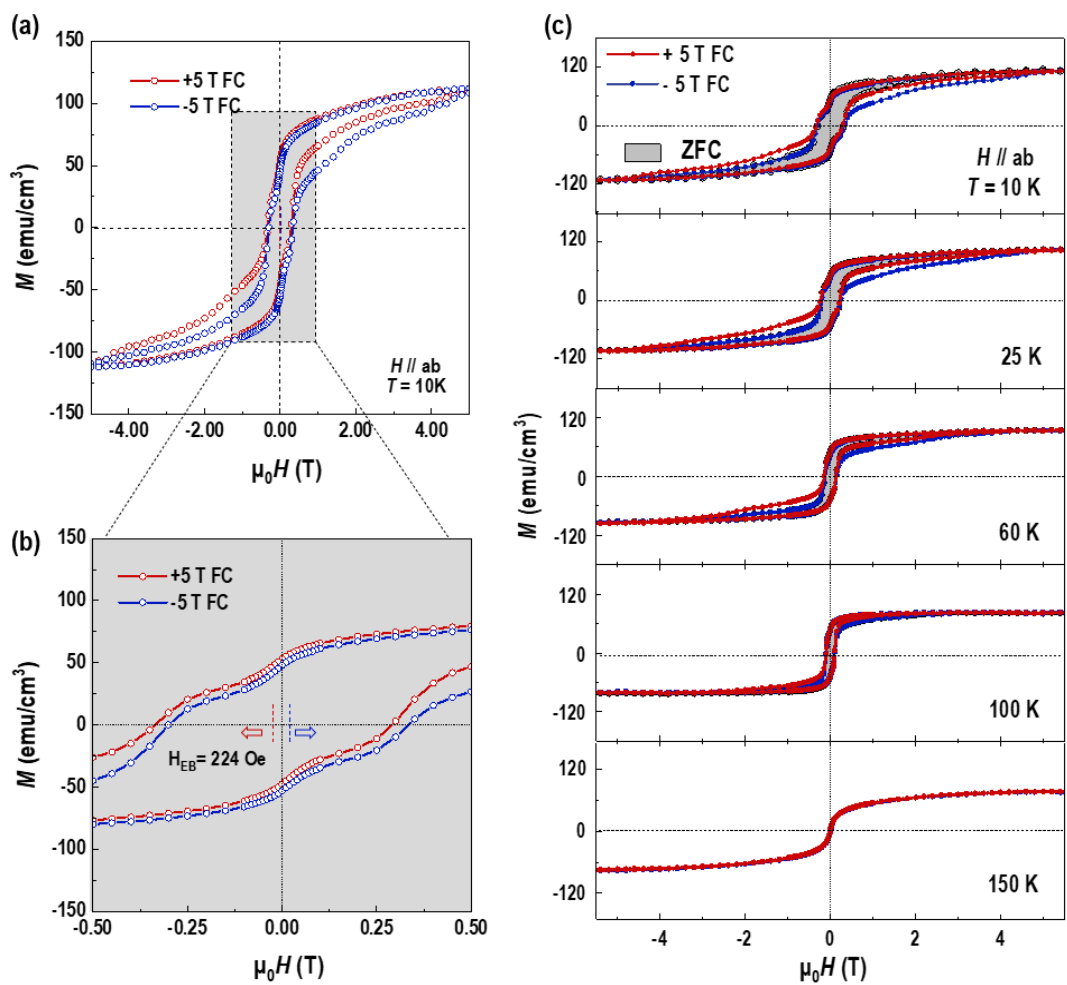

**Figure S3. Exchange bias effects observed at the RuO$_2$/SrRuO$_3$ interfaces.** (a) *M-H* hysteresis loops measured after field cooling of +5 T and -5 T at 10 K. The zoom-in M-H loops at the low field region is plotted in (b). (c) *M-H* hysteresis loops measured at various temperatures. As increasing temperature, the coercive field, saturation moment, and exchange bias effect reduce progressively.

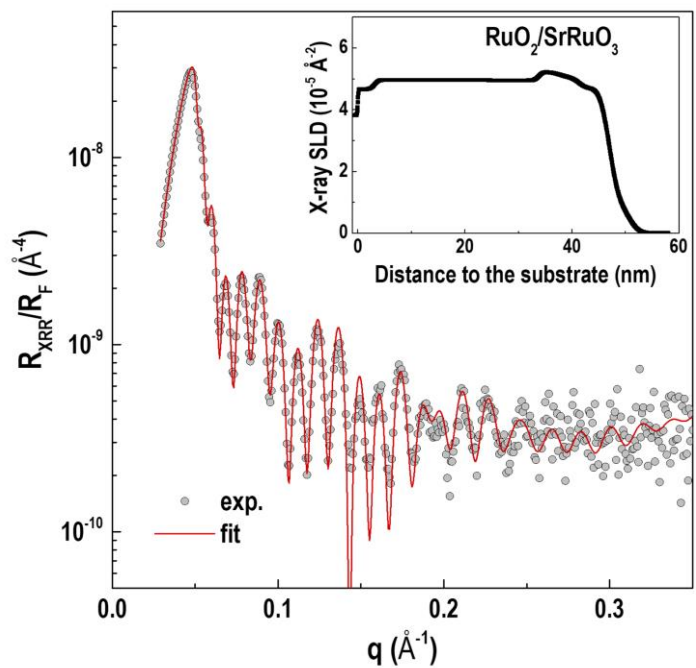

**Figure S4. X-ray reflectivity of the RuO₂/SrRuO₃ heterostructure.** The solid symbols and red lines represent experimental data and best fit, respectively. Inset shows the chemical depth profile across the RuO₂/SrRuO₃ heterostructure. The interfacial RuO₂ layer shows a slightly increased atomic density probably attributed to the structural modification.

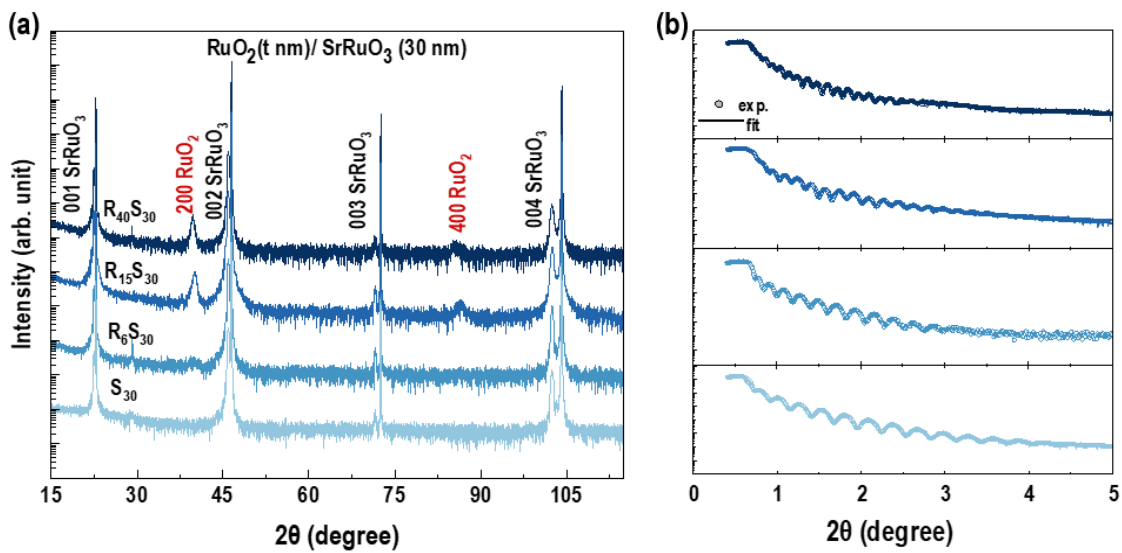

**Figure S5. Structural characterizations of $(RuO_2)_m/(SrRuO_3)_{30}$ ($R_mS_{30}$) heterostructures,** where $m$ and 30 represent the thickness (nm) of $RuO_2$ layer and $SrRuO_3$ layer, respectively. (a) XRD θ-2θ scans and (b) XRR curves of $R_mS_{30}$ heterostructures and $S_{30}$ single layer film. The open symbols and solid lines are experimental data and best fits, respectively.

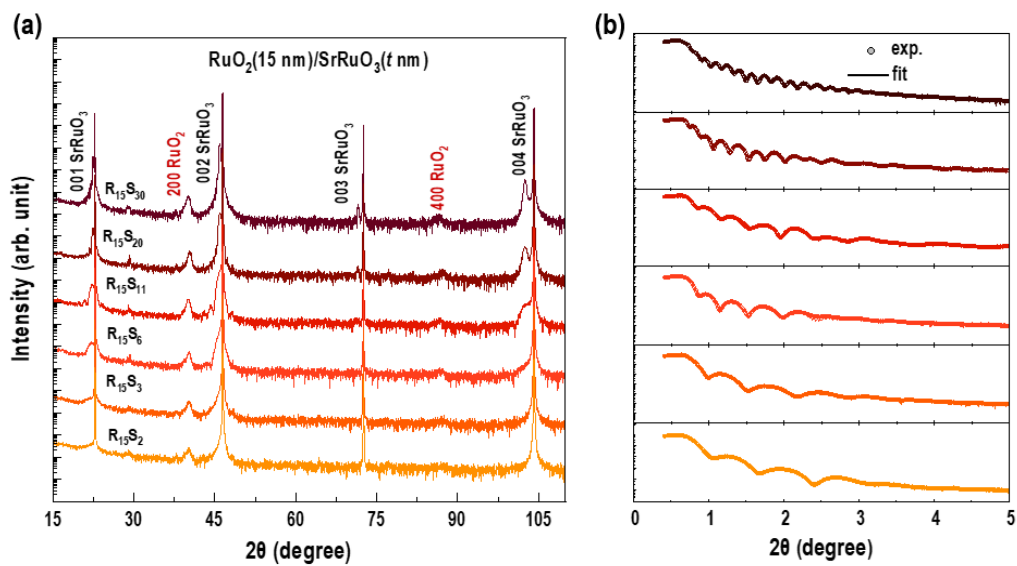

**Figure S6. Structural characterizations of [(RuO$_2$)$_{15}$/(SrRuO$_3$)$_n$] (R$_{15}$S$_n$) heterostructures**, where 15 and *n* represent the thickness (nm) of RuO$_2$ layer and SrRuO$_3$ layer, respectively. (a) XRD θ-2θ scans of R$_{15}$S$_n$ heterostructures with various *n*. (b) XRR curves of R$_{15}$S$_n$ heterostructures. The open symbols and solid lines are experimental data and best fits, respectively.

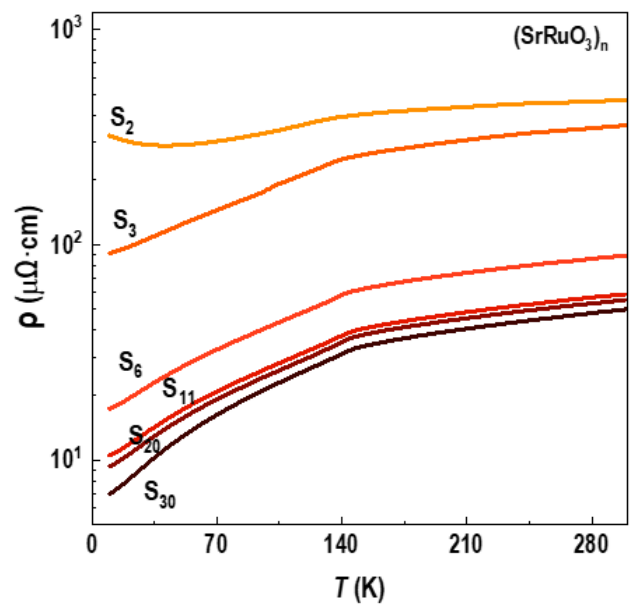

**Figure S7. Transport properties of SrRuO$_3$ (S$_n$) single layer films**, where *n* represent the thickness (nm) of SrRuO$_3$ layer. *R-T* curves of S$_n$ single layer films.

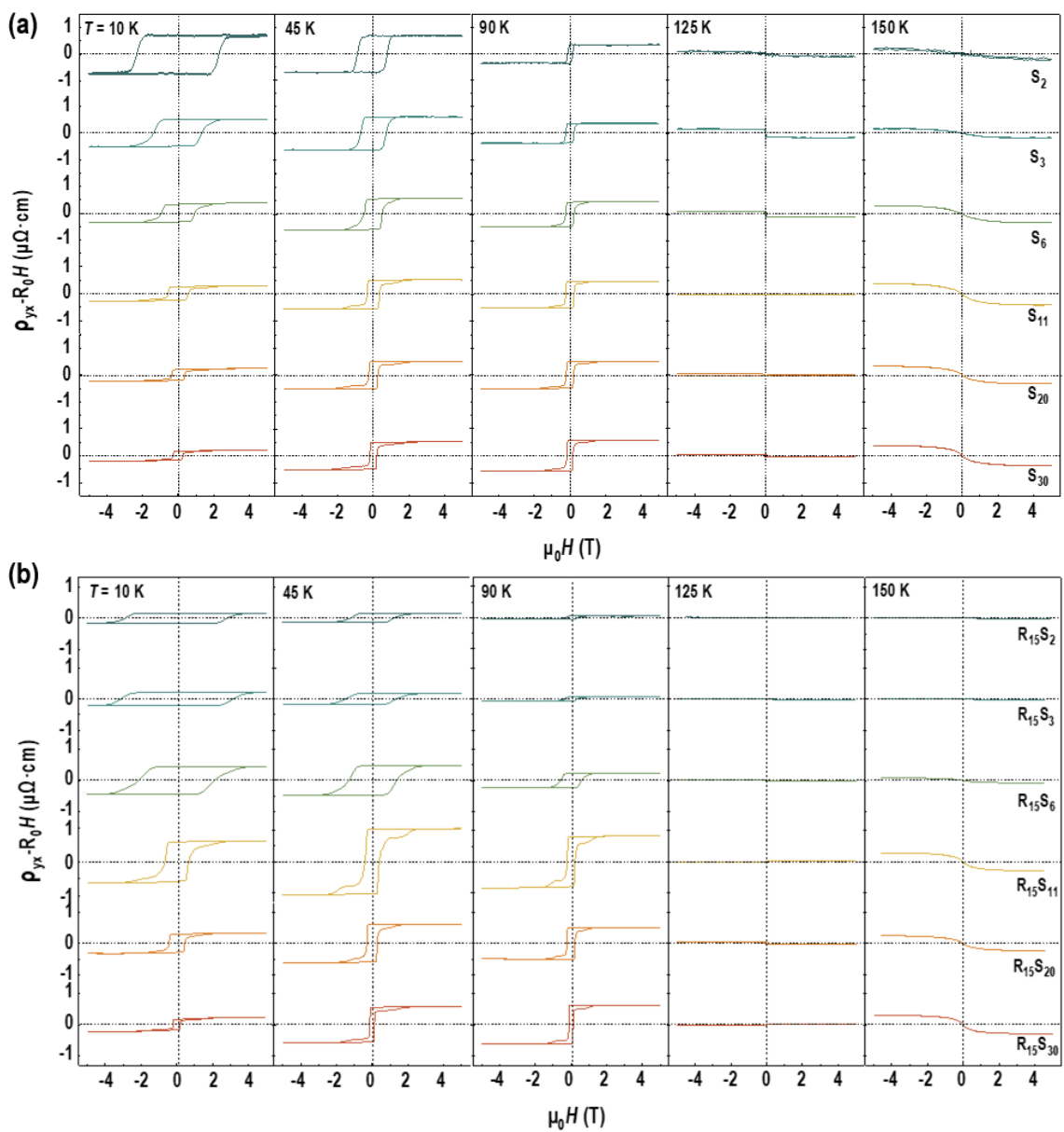

**Figure S8. Anomalous Hall resistivity** of (a) SrRuO$_3$ (S$_n$) single layer films and (b) (RuO$_2$)$_{15}$/(SrRuO$_3$)$_n$ (R$_{15}$S$_n$) heterostructures at different temperatures.

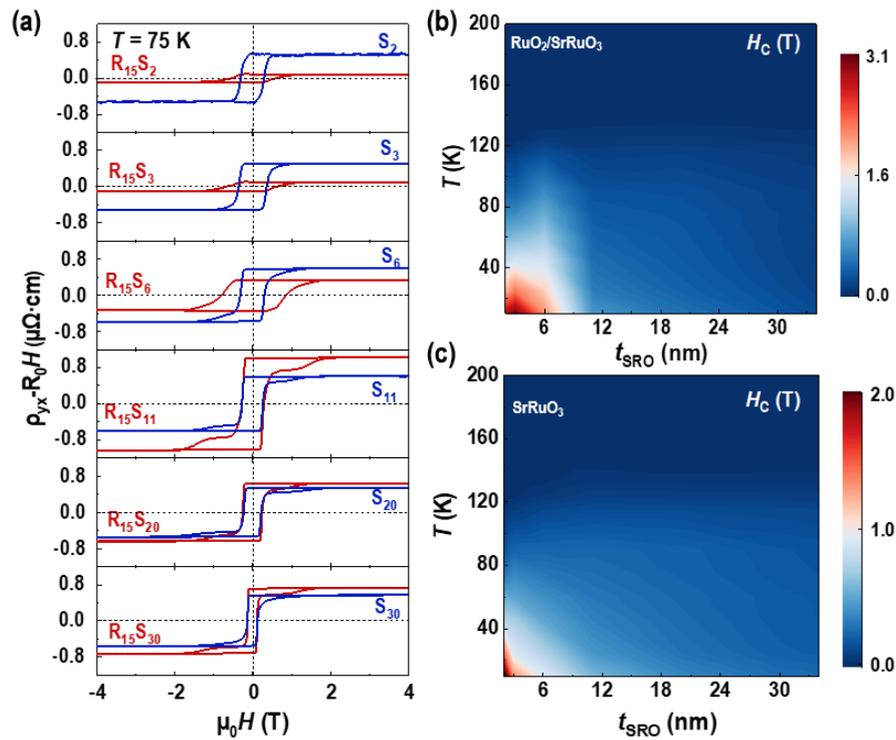

**Figure S9. Nonmonotonic AHE coefficient in the RuO₂/SrRuO₃ heterogeneous films with various thickness of SrRuO₃ layer.** (a) Magnetic field-dependent anomalous Hall resistivity ($\rho_{xy}-R_0H$) of [(RuO$_2$)$_{15}$/(SrRuO$_3$)$_n$] (R$_{15}$S$_n$) heterostructures and SrRuO$_3$ (S$_n$) single layer films at 75 K, where 15 and $n$ represent the layer thickness (nm) of RuO$_2$ and SrRuO$_3$ layer, respectively. Coercive field ($H_C$) as function of temperature and SrRuO$_3$ layer thickness ($t_{SRO}$) for RuO$_2$/SrRuO$_3$ heterostructures (b) and SrRuO$_3$ single layer films (c).

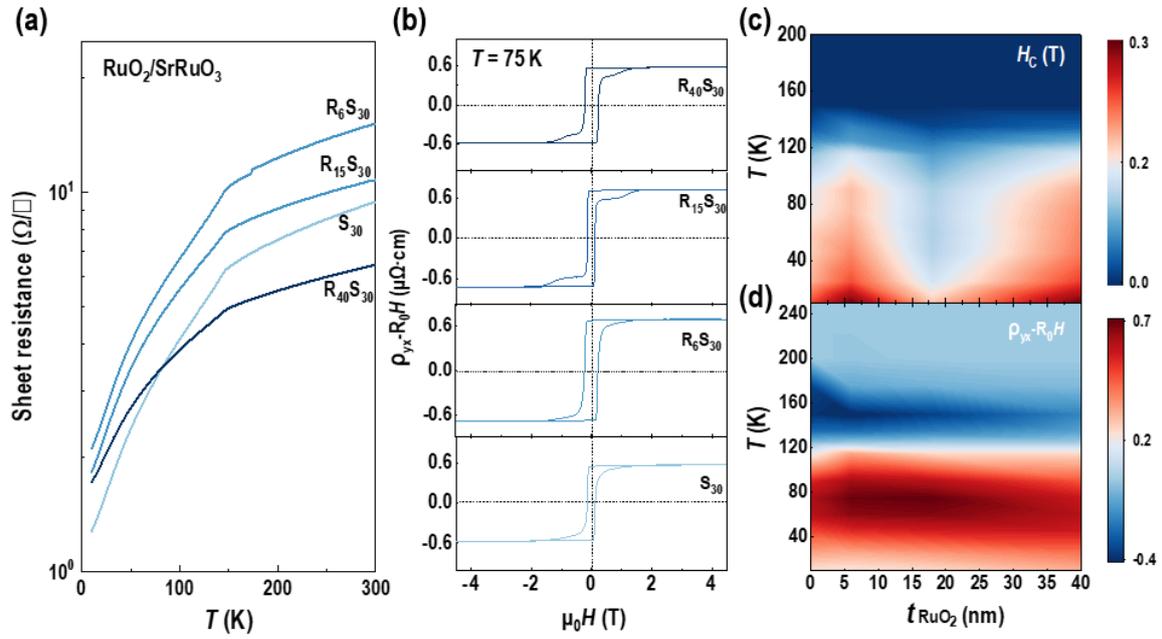

**Figure S10. Transport properties of (RuO$_2$)$_m$/(SrRuO$_3$)$_{30}$ (R$_m$S$_{30}$) heterostructures**, where *m* and 30 represent the thickness (nm) of RuO$_2$ layer and SrRuO$_3$ layer, respectively. (a) *R-T* curves of R$_m$S$_{30}$ heterostructures and S$_{30}$ single layer. (b) Field-dependent anomalous Hall resistivity $\rho_{xy} - R_0H$ of R$_m$S$_{30}$ heterostructures and S$_{30}$ single layer at *T* = 75 K. (c) Coercive field (*H*$_C$) and (f)$\rho_{xy} - R_0H$ of R$_m$S$_{30}$ heterostructures as function of temperature and RuO$_2$ layer's thickness.